\pdfoutput=1 
\documentclass[10pt,conference]{IEEEtran}

\usepackage[T1]{fontenc} 
\usepackage{amsmath,amssymb,amsfonts}
\usepackage{cite}
\usepackage{algorithmic}
\usepackage{graphicx}
\usepackage{textcomp}
\usepackage{xcolor}
\usepackage{multirow}
\usepackage{subcaption}
\usepackage{lipsum}

\usepackage[normalem]{ulem}

\usepackage{todonotes}
\usepackage{url}
\usepackage{graphicx}
\usepackage{xspace}
\usepackage{tikz,pgfplots}

\usepgfplotslibrary{statistics}

\newcommand{\system}{\textsc{Groot}\xspace}

\begin{document}

\title{Groot: An Event-graph-based Approach for Root Cause Analysis in Industrial Settings}

\author{
\IEEEauthorblockN{Hanzhang Wang\IEEEauthorrefmark{1}, Zhengkai Wu\IEEEauthorrefmark{2},
Huai Jiang\IEEEauthorrefmark{1},
Yichao Huang\IEEEauthorrefmark{1},
\\Jiamu Wang\IEEEauthorrefmark{1},
Selcuk Kopru\IEEEauthorrefmark{1},
Tao Xie\IEEEauthorrefmark{4}}
\IEEEauthorblockA{\IEEEauthorrefmark{1}eBay, \IEEEauthorrefmark{2}University of Illinois at Urbana-Champaign, \IEEEauthorrefmark{4}Peking University}

\IEEEauthorblockA{Email: \{hanzwang,huajiang,yichhuang,jiamuwang,skopru\}@ebay.com,
 zw3@illinois.edu,
taoxie@pku.edu.cn}
}

\maketitle

\begingroup\renewcommand\thefootnote{\textsection}
\footnotetext{Tao Xie is also affiliated with Key Laboratory of High Confidence Software Technologies (Peking University), Ministry of Education, China. Hanzhang Wang is the corresponding author.}
\endgroup

\begin{abstract}

For large-scale distributed systems, it is crucial to efficiently diagnose the root causes of incidents to maintain high system availability. The recent development of microservice architecture brings three major challenges (i.e., complexities of operation, system scale, and monitoring) to root cause analysis (RCA) in industrial settings. To tackle these challenges, in this paper, we present \system, an event-graph-based approach for RCA. \system constructs a real-time causality graph based on events that summarize various types of metrics, logs, and activities in the system under analysis. Moreover, to incorporate domain knowledge from site reliability engineering (SRE) engineers, \system can be customized with user-defined events and domain-specific rules. Currently, \system supports RCA among \textit{5,000} real production services and is actively used by the SRE teams in eBay, a global e-commerce system serving more than \textit{159 million} active buyers per year. Over 15 months, we collect a data set containing labeled root causes of 952 real production incidents for evaluation. The evaluation results show that \system is able to achieve 95\% top-3 accuracy and 78\% top-1 accuracy. To share our experience in deploying and adopting RCA in industrial settings, we conduct a survey to show that users of \system find it helpful and easy to use. We also share the lessons learned from deploying and adopting \system to solve RCA problems in production environments.

\end{abstract}


\section{Introduction}

\label{sec:intro}
Since the emergence of microservice  architecture~\cite{balalaie2016microservices}, it has been quickly adopted by many large companies such as Amazon, Google, and Microsoft.
Microservice architecture aims to improve the scalability, development agility, and reusability of these companies' business systems.
Despite these undeniable benefits, different levels of components in such a system can go wrong 
due to the fast-evolving and large-scale nature of microservices architecture~\cite{balalaie2016microservices}. 
Even if there are minimal human-induced faults in code, the system might still be at risk due to anomalies in hardware, configurations, etc. 
Therefore, it is critical to detect anomalies and then efficiently analyze the root causes of the associated incidents, subsequently helping the system reliability 
engineering (SRE) team take further actions to bring the system back to normal.

In the process of recovering a system, it is critical to conduct accurate and efficient root cause analysis (RCA)~\cite{sole2017survey}, the second one of a three-step process. In the first step, anomalies are detected with alerting mechanisms~\cite{zhao2020automatically, xu2017lightweight, tang2012optimizing} 
based on monitoring data such as logs~\cite{aguilera2003performance,zawawy2010log,nair2015learning,lu2017log,gan2019seer}, metrics/key performance indicators (KPIs)~\cite{mace2015pivot,xu2018unsupervised,ma2019ms, meng2020localizing, wu2020microrca}, or a combination thereof~\cite{kim2013root,wang2019grano}.  In the second step, when the alerts are triggered, RCA is performed to analyze the root cause of these alerts and additional events, and to propose recovery actions from the associated incident~\cite{baek2017cloudsight,da2016monitoring,aguilera2003performance}. RCA needs to consider multiple possible interpretations of potential causes for the incident, and these different interpretations could lead to different mitigation actions to be performed.  
In the last step, the SRE teams perform those mitigation actions and recover the system. 

Based on our industrial SRE experiences, we find that RCA is difficult in industrial practice due to three complexities, particularly under microservice settings:
\begin{itemize}
    \item \textbf{\emph{Operational Complexity}}. For large-scale systems, there are typically centered (aka infrastructure) SRE and domain (aka embedded) SRE engineers~\cite{GoogleSRE}. Their communication is often ineffective or limited under the microservice scenarios due to a more diversified tech stack, granular services, and shorter life cycles than traditional systems. The knowledge gap between the centered SRE team and the domain SRE team gets further enlarged and makes RCA much more challenging. Centered SRE engineers have to learn from domain SRE engineers on how the new domain changes work to update the centralized RCA tools. Thus, adaptive and customizable RCA is required instead of one-size-fits-all solutions. 
        
    \item \textbf{\emph{Scale Complexity}}. There could be thousands of services simultaneously running in a large microservice system, resulting in a very high number of monitoring signals. A real incident could cause numerous alerts to be triggered across services. The inter-dependencies and incident triaging between the services are proportionally more complicated than a traditional system~\cite{wu2020microrca}. To detect root causes that may be distributed and many steps away from an initially observed anomalous service, the RCA approach must be scalable and very efficient to digest high volume signals. 
    
    \item \textbf{\emph{Monitoring Complexity}}. A high quantity of observability data types (metrics, logs, and activities) need to be monitored, stored, and processed, such as intra-service and inter-service metrics. Different services in a system may produce different types of logs or metrics with different patterns. There are also various kinds of activities, such as code deployment or configuration changes. The RCA tools must be able to consume such highly diversified and unstructured data and make inferences. 


\end{itemize}

To overcome the limited effectiveness of existing approaches~\cite{sole2017survey,nguyen2013fchain,chen2014causeinfer,ma2020diagnosing,schoenfisch2018root,brandon2020graph,yoon2016dbsherlock,jeyakumar2019explainit,jayathilaka2017performance,zhao2020automatically,marvasti2013anomaly,weng2018root,qiu2020causality,meng2020localizing, kim2013root} (as mentioned in Section~\ref{sec:related}) in industrial settings due to the aforementioned complexities, we propose \system, an event-graph-based RCA approach. 
In particular, \system constructs an event causality graph, whose basic nodes are monitoring events such as performance-metric deviation events, status change events, and developer activity events. These events carry detailed information to enable accurate RCA. The events and the causalities between them are constructed using specified rules and heuristics (reflecting domain knowledge). In contrast to the existing fully learning-based approaches~\cite{gan2019seer, ma2020diagnosing, zhao2020automatically}, \system provides better transparency and interpretability. Such interpretability is critical in our industrial settings because a graph-based approach can offer visualized reasoning with causality links to the root cause and details of every event instead of just listing the results. Besides, our approach can enable effective tracking of cases and targeted detailed improvements, e.g., by enhancing the rules and heuristics used to construct the graph. 

\system has two salient advantages over existing graph-based approaches:
\begin{itemize}
\item \emph{\textbf{Fine granularity}} (events as basic nodes). First, unlike existing graph-based approaches, which directly use services~\cite{brandon2020graph} or hosts (VMs)~\cite{weng2018root} as basic nodes, \system constructs the causality graph by using monitoring events as basic nodes. Graphs based on events from the services can provide more accurate results to address the monitoring complexity. Second, for the scale complexity, \system can dynamically create hidden events or additional dependencies based on the context, such as adding dependencies to the external service providers and their issues. Third, to construct the causality graph, \system takes the detailed contextual information of each event into consideration for analysis with more depth. Doing so also helps \system incorporate SRE insights with the context details of each event to address the operational complexity.
\item \emph {\textbf{High diversity}} (a wide range of event types supported). First, the causality graph in \system supports various event types such as performance metrics, status logs, and developer activities to address the monitoring complexity. This multi-scenario graph schema can directly boost the RCA coverage and precision. For example, \system is able to detect a specific configuration change on a service as the root cause instead of performance anomaly symptoms, thus reducing triaging efforts and time-to-recovery (TTR). Second, \system allows the SRE engineers to introduce different event types that are powered by different detection strategies or from different sources. For the rules that decide causality between events, we design a grammar that allows easy and fast implementations of domain-specific rules, narrowing the knowledge gap of the operational complexity. Third, \system provides a robust and transparent ranking algorithm that can digest diverse events, improve accuracy, and produce results interpretable by visualization. 
\end{itemize}

To demonstrate the flexibility and effectiveness of \system, we evaluate it on eBay's production system that serves more than \textbf{159} million active users and features more than \textbf{5,000} services deployed over three data centers. 
We conduct experiments on a labeled and validated data set to show that \system achieves 95\% top-3 accuracy and 78\% top-1 accuracy for 952 real production incidents collected over 15 months. 
Furthermore, \system is deployed in production 
for real-time RCA, and is used daily by both centered and domain SRE teams, with the achievement of 73\% top-1 accuracy in action. 
Finally, the end-to-end execution time of \system for each incident in our experiments is less than 5 seconds, demonstrating the high efficiency of \system.

We report our experiences and lessons learned when using \system to perform RCA in the industrial e-commerce system. We survey among the SRE users and developers of \system, who find \system easy to use and helpful during the triage stage. Meanwhile, the developers also find the \system design to be desirable to make changes and facilitate new requirements. We also share the lessons learned from adopting \system in production for SRE in terms of technology transfer and adoption.

In summary, this paper makes four main contributions:
\begin{itemize}
    \item An event-graph-based approach named \system for root cause analysis tackling challenges in industrial settings.
    \item Implementation of \system in an RCA framework for allowing the SRE teams to instill domain knowledge.
    \item Evaluation performed in eBay's production environment with more than 5,000 services, for demonstrating \system's effectiveness and efficiency.
    \item Experiences and lessons learned when deploying and applying \system in production.
\end{itemize}

\section{Related Work}
\label{sec:related}


\emph{Anomaly Detection.} Anomaly detection aims to detect potential issues in the system. Anomaly detection approaches using time series data can generally be categorized into three types: (1) batch-processing and historical analysis such as  Surus~\cite{nflxsurus}; (2) machine-learning-based, such as Donut~\cite{xu2018unsupervised}; (3) usage of adaptive concept drift, such as StepWise~\cite{ma2018robust}.


\system currently uses a combination of manually written thresholds, statistical models, and machine learning (ML) algorithms to detect  anomalies. Since our approach is event-driven, as long as fairly accurate alerts are generated, \system is able to incorporate them. 


\emph{Root Cause Analysis.} Traditional RCA approaches (e.g., Adtributor~\cite{bhagwan2014adtributor} and HotSpot~\cite{sun2018hotspot}) find the multi-dimensional combination of attribute values that would lead to certain quality of service (QoS) anomalies. These approaches are effective at discrete static data. Once there are continuous data introduced by time series information, these approaches would be much less effective.

To tackle these difficulties, there are two categories of approaches based on ML and graph, respectively. 

\emph{ML-based RCA.} Some ML-based approaches use features such as time series information~\cite{ma2020diagnosing, weng2018root} and features extracted using textual and temporal information~\cite{zhao2020automatically}. Some other approaches~\cite{xu2018unsupervised} conduct deep learning by first constructing the dependency graph of the system and then representing the graph in a neural network. However, these ML-based approaches face the challenge of lacking training data. Gan et al.~\cite{gan2019seer} proposed Seer to make use of historical tracking data. Although Seer also focuses on the microservice scenario, it is designed to detect QoS violations while lacking support for other kinds of errors. There is also an effort to use unsupervised learning such as GAN~\cite{xu2018unsupervised}, but it is generally hard to simulate large, complicated distributed systems to give meaningful data.

\emph{Graph-based RCA.} A recent survey~\cite{sole2017survey} on RCA approaches categorizes more than 20 RCA algorithms by more than 10 theoretical models to represent the relationships between components in a microservice system.  
Nguyen et al.~\cite{nguyen2013fchain} proposed FChain, which introduces time series information into the graph, but they still use server/VM as nodes in the graph. Chen et al.~\cite{chen2014causeinfer}  proposed CauseInfer, which constructs a two-layered hierarchical causality graph. It applies metrics as nodes that indicate service-level dependency. Schoenfisch et al.~\cite{schoenfisch2018root} proposed to use Markov Logic Network to express conditional dependencies in the first-order logic, but still build dependency on the service level. Lin et al.~\cite{lin2018microscope} proposed Microscope, which targets the microservice scenario. It builds the graph only on service-level metrics so it cannot get full use of other information and lacks customization. Brandon et al. ~\cite{brandon2020graph} proposed to build the system graph using metrics, logs, and anomalies, and then use pattern matching against a library to identify the root cause. However, it is difficult to update the system to facilitate the changing requirements. Wu et al.~\cite{wu2020microrca} proposed MicroRCA, which models both services and machines in the graph and tracks the propagation among them. It would be hard to extend the graph from machines to the concept of other resources such as databases in our paper. 


As mentioned in Section~\ref{sec:intro}, by using the event graph, \system mainly overcomes the limitations of existing graph-based approaches in two aspects: (1) build a more accurate and precise causality graph use the event-graph-based model; (2) allow adaptive customization of link construction rules to incorporate domain knowledge in order to facilitate the rapid requirement changes in the microservice scenario.

Our \system approach uses a customized page rank algorithm in the event ranking, and can also be seen as an unsupervised ML approach. Therefore, \system is complementary to other ML approaches as long as they can accept our event causality graph as a feature.

\begin{table}[]
\centering
\caption{The scale of experiments in existing RCA approaches’ evaluations (QPS: Queries per second)}
\resizebox{0.45\textwidth}{!}{ 
\begin{tabular}{|l|l|l|l|}
\hline
Approach      & Year   & Scale    & Validated on Real Incidents?                             \\ \hline
FChain~\cite{nguyen2013fchain}        & 2013                  & \textless{}= 10 VMs         & No           \\ \hline
CauseInfer~\cite{chen2014causeinfer}    & 2014             & 20 services on 5 servers       & No       \\ \hline
MicroScope~\cite{lin2018microscope}    & 2018                       & 36 services, $\sim$5000 QPS & No          \\ \hline
APG~\cite{weng2018root}           & 2018                  & \textless{}=20 services on 5 VMs   & No   \\ \hline
Seer~\cite{gan2019seer}          & 2019   & \textless{}=50 services on 20 servers & Partially \\ \hline
MicroRCA~\cite{wu2020microrca}      & 2020                        & 13 services, $\sim$600 QPS     & No       \\ \hline
RCA Graph~\cite{brandon2020graph}     & 2020                          & \textless{}=70 services on 8 VMs  & No    \\ \hline
Causality RCA~\cite{qiu2020causality} & 2020                     & \textless{}=20 services        & No       \\ \hline
\end{tabular}
}
\label{tab:related}
\end{table}


\emph{Settings and Scale.} The challenges of operational, scale, and monitoring complexities are observed, especially being substantial in the industrial settings. Hence, we believe that the target RCA approach should be validated at the enterprise scale and against actual incidents for effectiveness. 

Table~\ref{tab:related}  lists  the experimental  settings and scale in existing RCA approaches' evaluations.  All the listed existing approaches are evaluated in a relatively small scenario. 
In contrast, our experiments are performed upon a system containing 5,000 production services on hundreds of thousands of VMs. On average, the sub-dependency graph (constructed in Section~\ref{sec:appgraph}) of our service-based data set is already 77.5 services, more than the total number in any of the listed evaluations. Moreover, 7 out of the 8 listed approaches are evaluated under simulative fault injection on top of existing benchmarks such as RUBiS, which cannot represent real-world incidents;  Seer~\cite{gan2019seer}  collects only the real-world results with no validations. Our data set contains 952 actual incidents collected from real-world settings.
\section{Motivating Examples}

\label{sec:example}

In this section, we demonstrate the effectiveness of event-based graph and adaptive customization strategies with two motivating examples.

\begin{figure*}[t]
\begin{subfigure}[b]{0.37\textwidth}
\centering
  \includegraphics[width=0.98\textwidth]{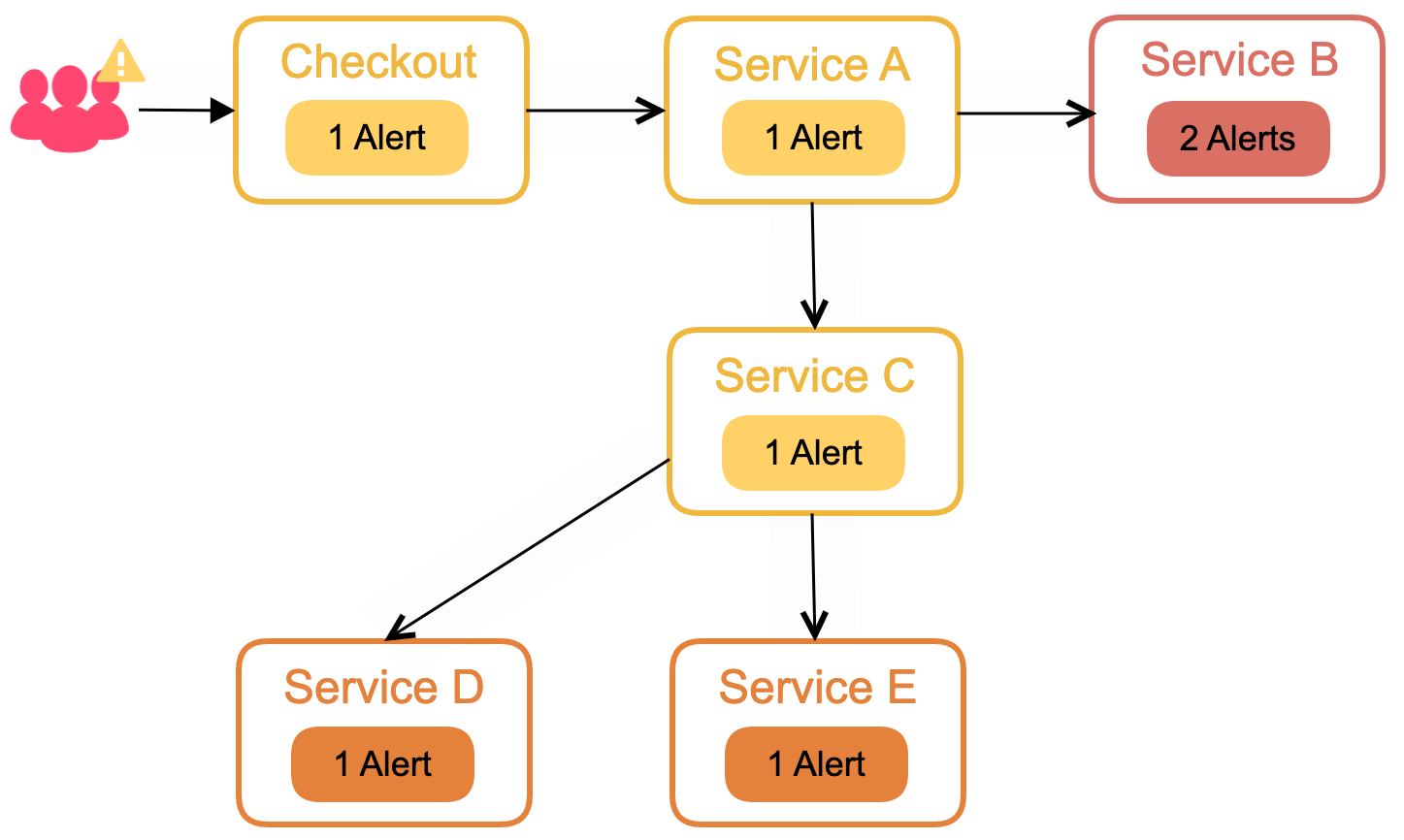}
  \caption{Dependency graph}
  \label{fig:ex1_dep}
\end{subfigure}
\hfill 
\begin{subfigure}[b]{0.4\textwidth}
\centering
  \includegraphics[width=0.98\textwidth]{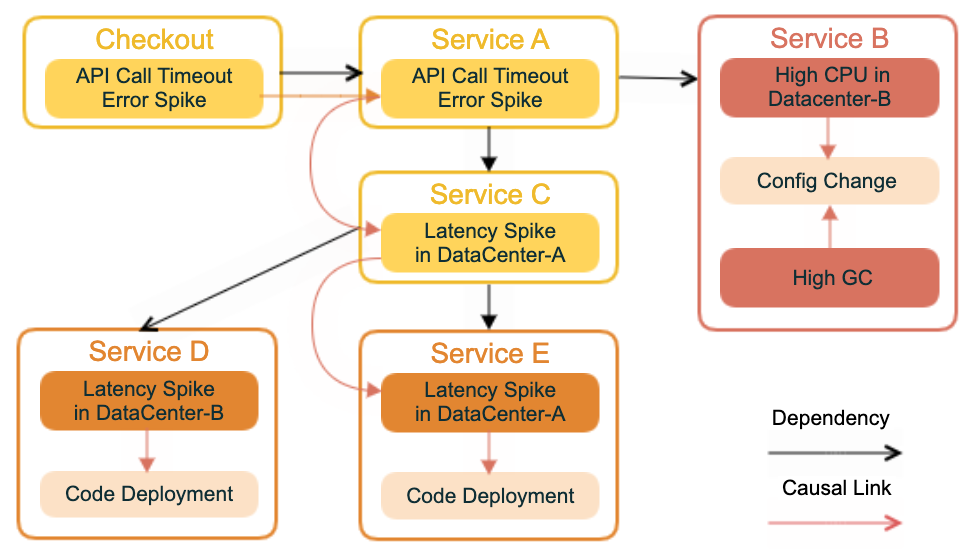}
  \caption{Causality graph}
  \label{fig:ex1_cas}
\end{subfigure}
\caption{Motivating example of event causality graph}
\label{fig:example1}
\end{figure*}

Figure~\ref{fig:example1} shows an abstracted real incident example with the dependency graph and the corresponding causality graph constructed by \system. The \emph{Checkout} service of our e-commerce system suddenly gets an additional latency spike due to a code deployment on the \emph{Service-E}. The service monitor is reporting \emph{API Call Timeout} detected by the ML-based anomaly detection system. The simplified sub-dependency graph consisting of 6 services is shown in Figure~\ref{fig:ex1_dep}. The initial alert is triggered on the \emph{Checkout} (entrance) service. The other nodes \emph{Service-*} are the internal services that the \emph{Checkout} service directly or indirectly depends on. The color of the nodes in Figure~\ref{fig:ex1_dep} indicates the severity/count of anomalies (alerts) reported on each service. We can see that \emph{Service-B} is the most severe one as there are two related alerts on it. The traditional graph-based approach~\cite{brandon2020graph,weng2018root} usually takes into account only the graph between services in addition to the severity information on each service. If the traditional approach got applied on Figure~\ref{fig:ex1_dep}, either \emph{Service-B}, \emph{Service-D}, or \emph{Service-E} could be a potential root cause, and \emph{Service-B} would have the highest possibility since it has two related alerts. Such results are not useful to the SRE teams. 

\system constructs the event-based causality graph as shown in Figure~\ref{fig:ex1_cas}. The events in each service are used as the nodes here. 
We can see that the \emph{API Call Timeout} issue in \emph{Checkout} is possibly caused by \emph{API Call Timeout} in \emph{Service-A}, which is further caused by \emph{Latency Spike} in \emph{DataCenter-A} of \emph{Service-C}. \system further tracks back to find that it is likely caused by \emph{Latency Spike} in \emph{Service-E}, which happens in the same data center. Finally \system figures out that the most probable root cause is a recent \emph{Code Deployment} event in \emph{Service-E}. The SRE teams then could quickly locate the root cause and roll back this code deployment, followed by further investigations.

There are no casual links between events in \emph{Service-B} and \emph{Service-A}, since no causal rules are matched. The \emph{API Call Timeout} event is less likely to depend on the event type \emph{High CPU} and \emph{High GC}. Therefore, the inference can eliminate \emph{Service-B} from possible root causes. This elimination shows the benefit of the event-based graph. Note that there is another event \emph{Latency Spike} in \emph{Service-D}, but not connected to \emph{Latency Spike} in \emph{Service-C} in the causality graph. The reason is that the \emph{Latency Spike} event in \emph{Service-C} happens in \emph{DataCenter-A}, not \emph{DataCenter-B}. 

\begin{figure}[t]
\centering
\resizebox{0.4\textwidth}{!}{ 
  \includegraphics[width=0.44\textwidth]{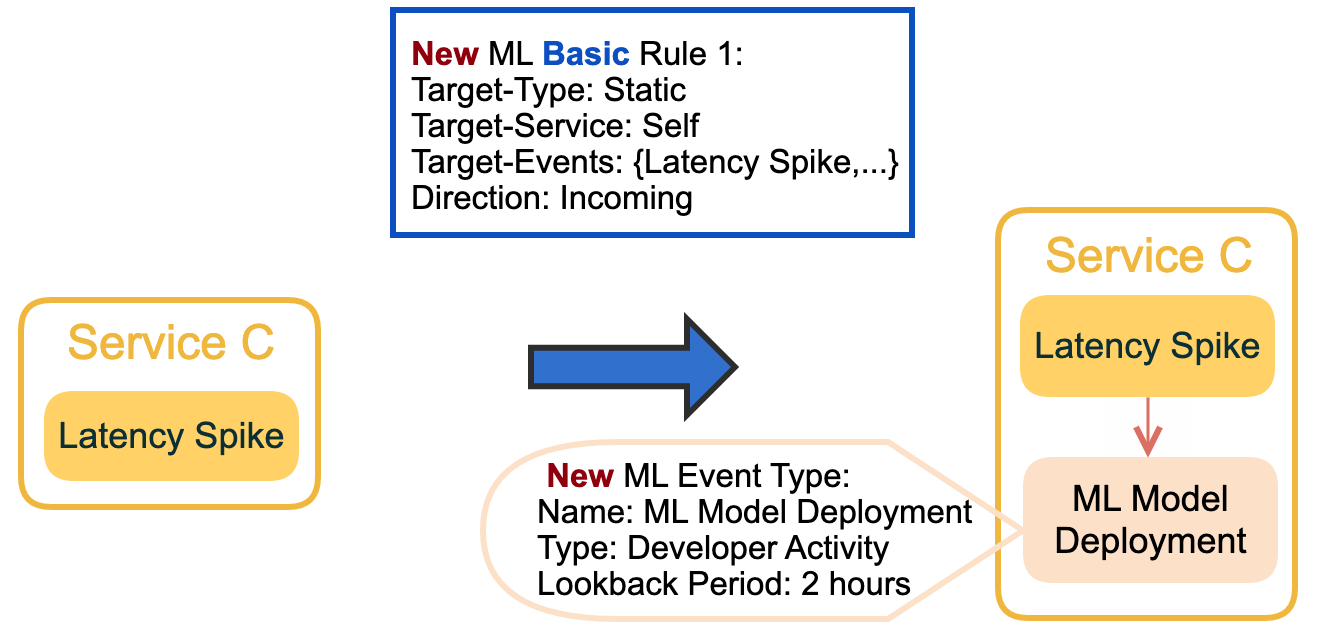}
  }
  \caption{Example of event type addition}
  \label{fig:ex2_n1}
\end{figure}

\begin{figure}[t]
\centering
\resizebox{0.44\textwidth}{!}{ 
  \includegraphics[width=0.44\textwidth]{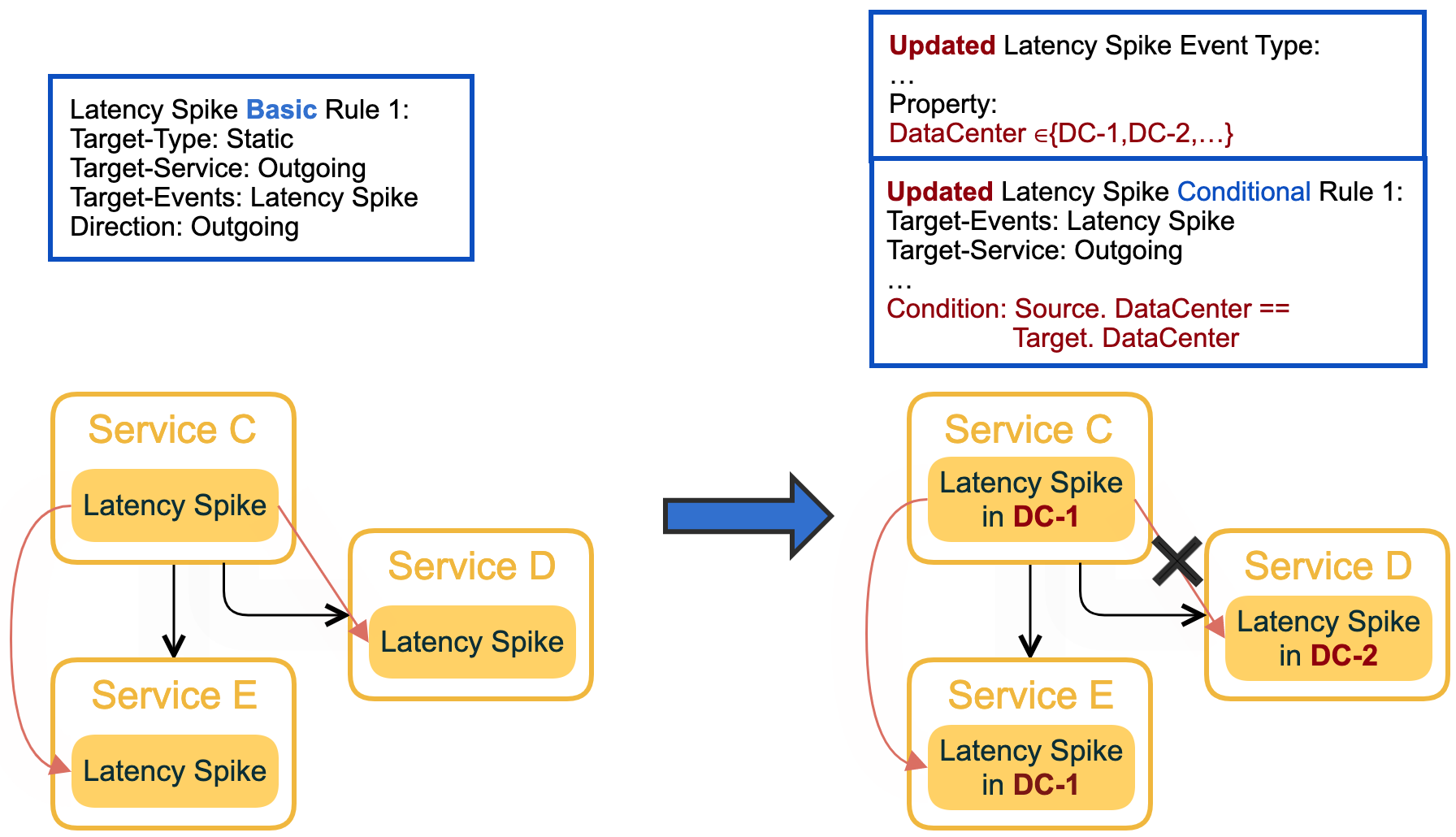}
}
  \caption{Example of event and rule update}
  \label{fig:ex2_n2}
\end{figure}

Figures~\ref{fig:ex2_n1} and~\ref{fig:ex2_n2} show how SRE engineers can easily change \system to adapt to new requirements, by updating the events and rules. In Figure~\ref{fig:ex2_n1}, SRE engineers want to add a new type of deployment activity, \emph{ML Model Deployment}. Usually, the SRE engineers first need to select the anomaly detection model or set their own alerts and provide alert/activity data sources for the stored events. In this example, the event can be directly fetched from the ML model management system. Then \system also requires related properties (e.g., the detection time range) to be set for the new event type. Lastly, the SRE engineers add the rules for building the causal links between the new event type and existing ones. The blue box in Figure~\ref{fig:ex2_n1} shows the rule, which denotes the edge direction, target event, and target service (self, upstream, and downstream dependency). 

Figure~\ref{fig:ex2_n2} shows a real-world example of how \system is able to incorporate SRE insights and knowledge. More specifically, SRE engineers would like to change the rules to allow \system to distinguish the latency spikes from different data centers. As an example in Figure \ref{fig:ex1_cas}, \emph{Latency Spike} events propagate only within the same data center. During \system development, SRE engineers could easily add new property \emph{DataCenter} to the \emph{Latency Spike} event. Then they add the corresponding ``conditional'' rules to be differentiated with the ``basic'' rules in Figure~\ref{fig:ex2_n2}. In conditional rules, links are constructed only when the specified conditions are satisfied. 
\section{Approach}
\begin{figure}[t]
\centering
\resizebox{0.48\textwidth}{!}{ 
  \includegraphics[width=\textwidth]{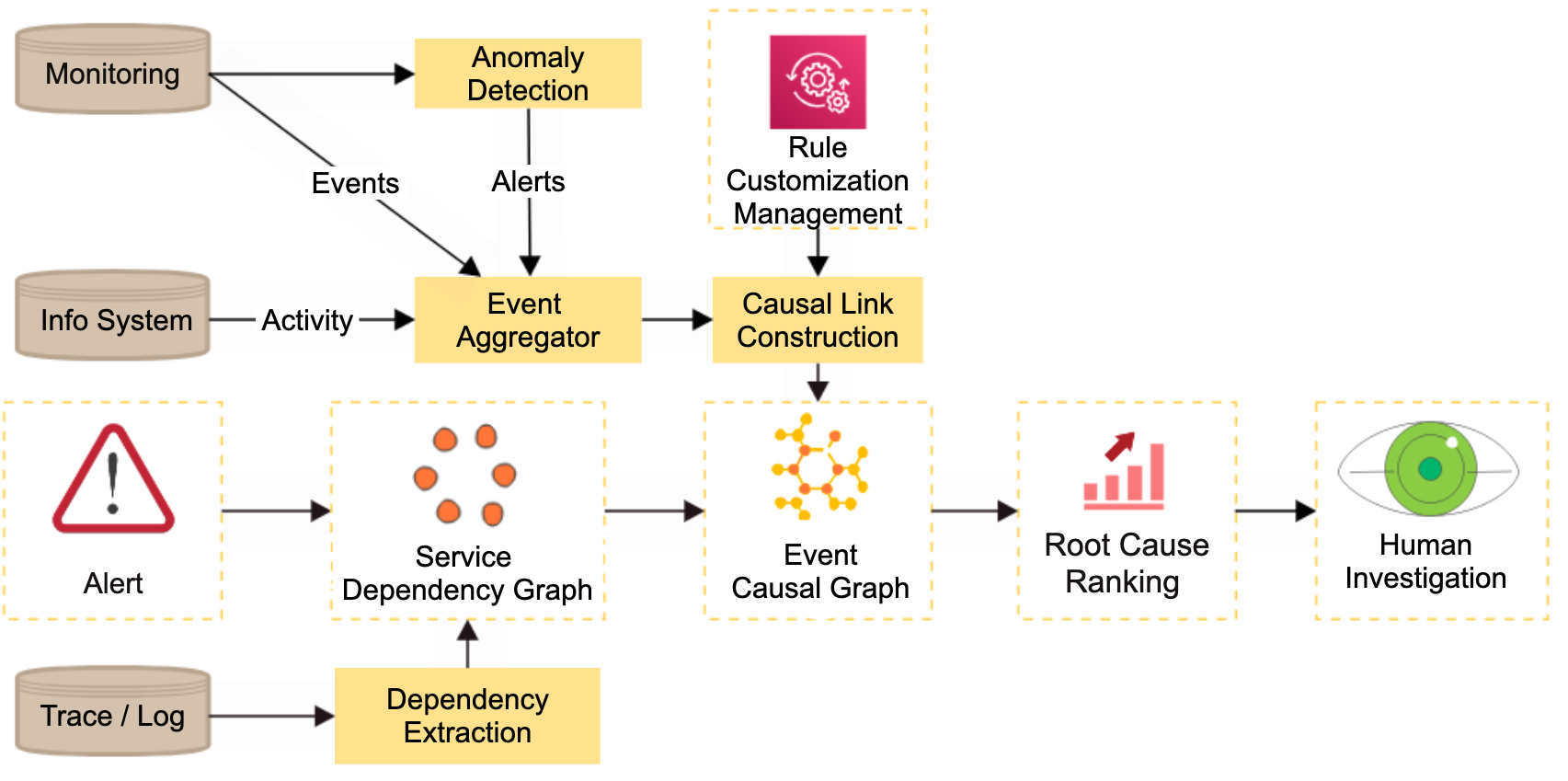}
}
  \caption{Workflow of \system}
  \label{fig:workflow}
\end{figure}

Figure ~\ref{fig:workflow} shows the overall workflow of \system. The triggers for using \system are usually alert(s) from automated anomaly detection, or sometimes an SRE engineer's suspicion. There are three major steps: constructing the service  dependency graph, constructing the event causality graph,  and root cause ranking. The outputs are the root causes ranked by the likelihood. To support fast human investigation experience, we build an interactive UI as shown in  Figure~\ref{fig:UI}: the service dependency, events with causal links and additional details such as raw metrics or the developer contact (of a code deployment event) are presented to the user for next steps. As an  offline part of human investigation, we label/collect a data set, perform validation, and summarize the knowledge for further improvement on all incidents on a daily basis. 

\subsection{Constructing Service Dependency Graph}
\label{sec:appgraph}

The construction of the service dependency graph starts with the initial alerted or suspicious service(s), denoted as $I$. For example, in Figure ~\ref{fig:ex1_dep}, $I=\{\textit{Checkout}\}$. $I$ can contain multiple services based on the range of the trigger alerts or suspicions. We maintain domain service lists where domain-level alerts can be triggered because there is no clear service-level indication.

At the back end, \system maintains a global service dependency graph $G_{global}$ via distributed tracing and log analysis. The directed edge from nodes $A$ to $B$ (two services or system components) in the dependency graph indicates a service invocation or other forms of dependency. In Figure~\ref{fig:ex1_dep}, the black arrows indicate such edges. Bi-directional edges and cycles between the services can be possible and exist. In this work, the global dependency graph is updated daily.

The service dependency (sub)graph $G$ is constructed using $G_{global}$ and $I$. An extended service list $L$ is first constructed by traversing each service in $I$ over $G_{global}$ for a radius range $r$. Each service $u \in L$ can be traversed by at least one service $v \in I$ within $r$ steps: $L=\{u|\exists v\in I, \ dist(u,v)\le r\ or\ dist(v,u)\le r\}$. Then, the service dependency subgraph $G$ is constructed by the nodes in $L$ and the edges between them in $G_{global}$. In our current implementation, $r$ is set to $2$, since this dependency graph may be dynamically extended in the next steps based on events' detail for longer issue chains or additional dependencies.

\subsection{Constructing Event Causality Graph}
\label{sec:causality}

In the second step, \system collects all supported events for each service in $G$ and constructs the causal links between events. 

\subsubsection{Collecting Events}

Table~\ref{tab:events} presents some example event types and detection techniques for \system's production implementation. For detection techniques, ``De Facto'' indicates that the event can be directly collected via a specific API or storage. 
The detection either runs passively in the back end to reduce delay and improve accuracy, or runs actively for only the services within the dependency graph range to save resources. 

There are three major categories of events: performance metrics, status logs, and developer activities:
\begin{itemize}
    \item \emph{Performance metrics} represent an anomaly of monitored time series metrics. For example, high CPU usage indicates that the service is causing high CPU usage on a certain machine. In this category, most events are continuously and passively detected and stored. 
    \item \emph{Status logs} are caused by abnormal system status, such as spike of HTTP error code metrics while accessing other services' endpoints. Different types of error metrics are important and supported in \system, including third-party APIs. For example, Bad Host indicates abnormal patterns on some machines running the service, and can be detected by a  clustering-based ML approach.
    \item \emph{Developer activities} are the events generated when a certain activity of developers is triggered, such as code deployment and config change.
\end{itemize}

\begin{table}[t]
\centering
\caption{List of example event types used in \system}
\resizebox{0.4\textwidth}{!}{ 
\begin{tabular}{|c|c|c|}
\hline
Type                                & Event Type                  & Detection Technique  \\ \hline
\multirow{6}{*}{Performance Metrics} & High GC (Overhead)      & Rule-based        \\ \cline{2-3} 
                                    & High CPU Usage          & Rule-based        \\ \cline{2-3} 
                                    & Latency Spike           & Statistical Model \\ \cline{2-3} 
                                    & TPS Spike               & Statistical Model \\ \cline{2-3} 
                                    & Database Anomaly        & ML Model          \\ \cline{2-3} 
                                    & Business Metric Anomaly & ML Model          \\ \hline
\multirow{4}{*}{Status Logs}        & WebAPI Error            & Statistical Model \\ \cline{2-3} 
                                    & Internal Error          & Statistical Model \\ \cline{2-3} 
                                    & ServiceClient Error     & Statistical Model \\ \cline{2-3} 
                                    & Bad Host                & ML Model          \\ \hline 
\multirow{3}{*}{Developer Activities} & Code Deployment         & De Facto          \\ \cline{2-3} 
                                    & Configuration Change    & De Facto          \\ \cline{2-3} 
                                    & Execute URL             & De Facto          \\ \hline
\end{tabular}
}
\label{tab:events}
\end{table}

In Groot, there are more than a dozen event types such as \emph{Latency Spike} as listed in the column 2 of Table~\ref{tab:events}. 
Each event type is characterized by three aspects: $Name$ indicates the name of this event type; $Lookback Period$ 
indicates the time range to look back (from the time when the use of \system is triggered) for collecting events of this event type;  $PropertyType$ indicates the types of the properties that an event of this event type should hold. 
$PropertType$  is characterized by a vector of pairs, each of which indicates the string type for a property's name and the primitive type for the property's value such as string, integer, and float. 
Formally, an event type is defined as a tuple: 
$ET = <Name, Lookback Period, PropertyType>$ 
where 
$PropertyType = <(string, \textit{type}_1), ..., (string, \textit{type}_{n})>$ ($n$ is the number of properties that an event of this event type holds). 

Each event of a certain event type $ET$ is characterized by four aspects:
$\textit{Service}$ indicates the service name that the event belongs to; $\textit{Type}$ indicates $ET$'s $\textit{Name}$;  $\textit{StartTime}$ indicates the time when the event happens; $\textit{Properties}$ indicates the properties that the event  holds.
Formally, an event is defined as a tuple: 
$e = <Service, Type, StartTime, Properties>$ 
where $Properties$ is an instantiation of $ET$'s  $PropertyType$.

%

For example, in Figure~\ref{fig:example1}, the generated event for \emph{Latency Spike in DataCenter-A} in \emph{Service-C} would be $<``\textit{Service-C}'', ``\textit{Latency\ Spike}'', \textit{2021/08/01-12:36:04}, <(``\textit{DataCenter}'',``\textit{DC-1}''),  ...>>$. 

\subsubsection{Constructing Causal Link}

After collecting all events on all services in $G$, in this step, causal links between these events are constructed for RCA ranking. The causal links (red arrows) in Figure~\ref{fig:ex1_cas} are such examples. A causal link represents that the source event can possibly be caused by the target event. SRE knowledge is engineered into rules and used to create causal links between the pairs of events. 

\begin{figure}[t]
\centering
\includegraphics[width=0.56\columnwidth]{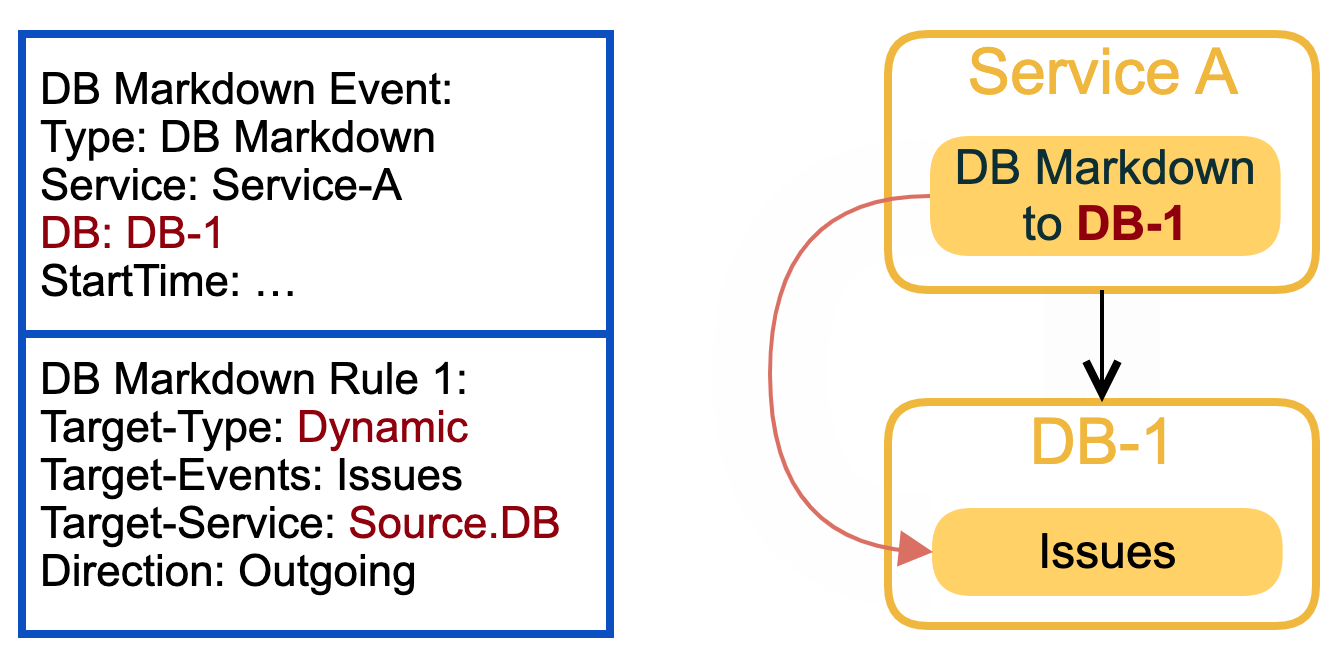}
\caption{Example of dynamic rule}
\label{fig:dynamic_example}
\end{figure}

A rule for constructing a causal link is defined as a tuple:  $Rule = <Target\mbox{-}Type,  Source\mbox{-}Events, Target\mbox{-}Events, Direction,\\ Target\mbox{-}Service,  Condition>$  ($Condition$ can be optionally specified). $Target\mbox{-}Type$ indicates the type of the rule, being either $Static$ or $Dynamic$ (explained further later). $Source\mbox{-}Events$ indicates the type of the causal link's source event ($Source\mbox{-}Events$ are listed in the names of the rules shown in Figures~\ref{fig:ex2_n1},~\ref{fig:ex2_n2} and~\ref{fig:dynamic_example}).   $Target\mbox{-}Events$ indicates the type of the causal link's target event. $Direction$ indicates the direction of the casual link between the target event and source event. $Target\mbox{-}Service$ indicates the service that the target event should belong to. Note that $Target\mbox{-}Service$ in $Static$ rules can be  $Self$, which indicates that the target event would be within the same service as the source event, or $Outgoing$/$Incoming$, which indicates that the target event would belong to the downstream/upstream services of the service that the source event belongs to in $G$.

There are two categories of special rules. The first category is \emph{dynamic} rules (i.e., rules whose $Target\mbox{-}Type$  is set to $Dynamic$) to support dynamic dependencies. Here $Target\mbox{-}Service$ does not indicate any of the three possible options listed earlier but indicates the name of the target service that \system would need to create. For example, live DB dependencies are not available due to different tech stacks and high volume. In Figure~\ref{fig:dynamic_example}, a DB issue (DB Markdown) is shown in \emph{Service-A}. Based on the listed \emph{dynamic} rule, \system creates a new ``service'' \emph{DB-1} in $G$, a new event ``Issues'' that belongs to \emph{DB-1}, and a causal link between the two events.  In practice, the SRE teams use dynamic rules to cover a lot of third-party services and database issues since the live dependencies are not easy to maintain.  

The second category of special rules is \emph{conditional} rules. \emph{Conditional} rules are used when some prerequisite conditions should be satisfied before a certain causal link is created. In these rules, $Condition$ is specified with a boolean predicate. As shown in Figure~\ref{fig:ex2_n2}, the SRE teams believe \emph{Latency Spike} events from different services are related only when both events happen within the same data center. Based on this observation, \system would first evaluate the predicate in $Condition$ and build only the causal link when the predicate is true. A conditional rule overwrites the basic rule on the same source-target event pair.

When constructing causal links, \system first applies the \emph{dynamic} rules so that dynamic dependencies and events are first created at once. Then for every event in the initial services (denoted as $I$), if the rule conditions are satisfied, one or many causal links are created from this event to other events from the same or upstream/downstream services. When a causal link is created, the step is repeated recursively for the target event (as a new origin) to create new causal links. After no new causal links are created, the construction of the event causality graph is finished.



\subsection{Root Cause Ranking}
Finally, \system ranks and recommends the most probable root causes from the event causality graph. Similar to how search engines infer the importance of pages by page links, we customize the PageRank \cite{manning2010introduction} algorithm to calculate the root cause ranking; the customized algorithm is named as GrootRank. The input is the event causality graph from the previous step. Each edge is associated with a weighted score for weighted propagation. The default value is set as $1$, and is set lower for alerts with high false-positive rates. 

Based on the observation that dangling nodes are more likely to be the root cause, we customize the personalization vector as $P_n = f_n $ or $P_d = 1$, where $P_d$ is the personalization score for dangling nodes, and $P_n$ is for the remaining nodes; and $f_n$ is a value smaller than 1 to enhance the propagation between dangling nodes. In our work, the parameter setting is $f_n = 0.5$, $\alpha = 0.85$, $max_{iter} = 100$ (which are parameters for the PageRank algorithm). Figure \ref{fig:person} illustrates an example. The grey circles are the events collected from three services and one database. The grey arrows are the dependency links and the red ones are the causal links with the weight of $1$. Both of the PageRank and GrootRank algorithms detect $event 5$ (DB issue) as the root cause, which is expected and correct. However, the PageRank algorithm ranks $event 4$ higher than $event 3$. But $event 3$ of $\textit{Service-C}$ is more likely to be the second most possible root cause (besides $event 5$), because the scores on dangling nodes are propagated to all others equally in each iteration. We can see that $event 3$ is correctly ranked as second using the GrootRank algorithm.

The second step of GrootRank is to break the tied results from the previous step. The tied results are due to the fact that the event graph can contain multiple disconnected sub-graphs with the same shape. We design two techniques to untie the ranking: 
\begin{figure}[t]
\centering
  \includegraphics[width=0.8\columnwidth]{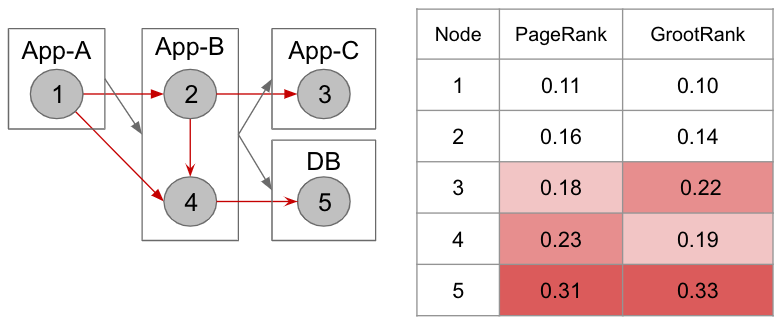}
  \caption{Example of personalization vector customization}
  \label{fig:person}
\end{figure}

\begin{figure}[t]
\centering
  \includegraphics[width=0.8\columnwidth]{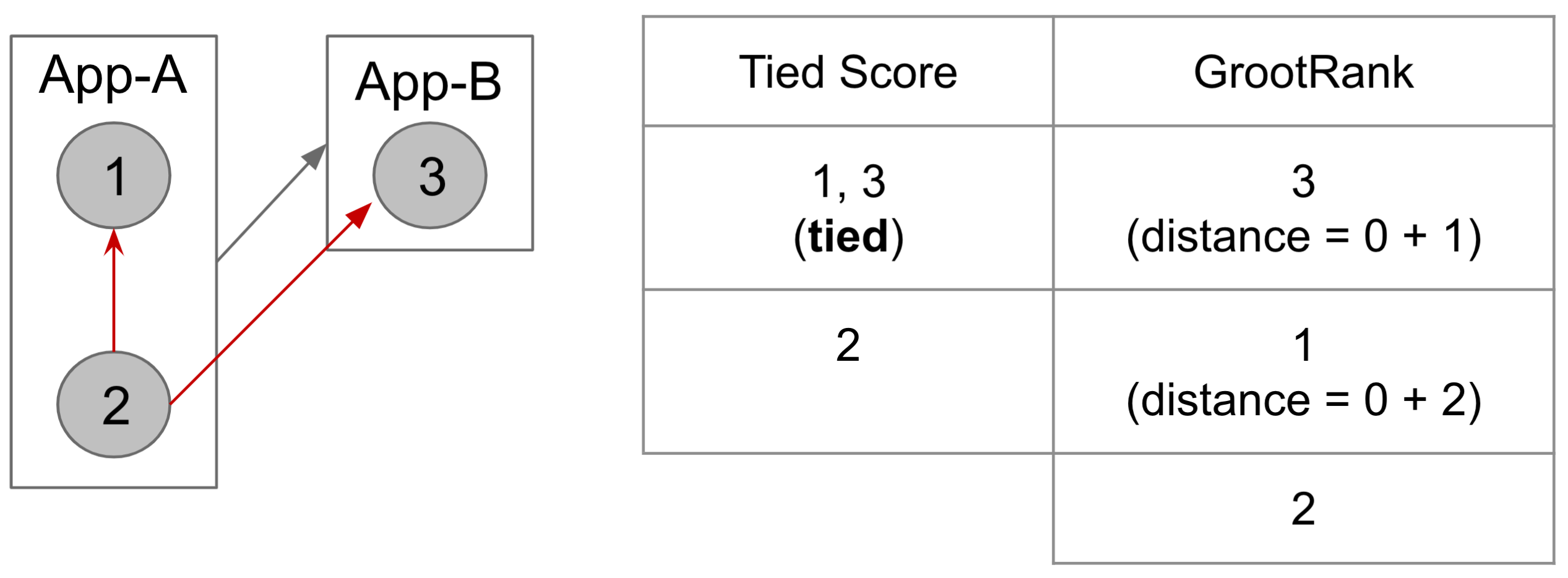}
  \caption{Example of using access distance to untie the ranking results}
  \label{fig:untie}
\end{figure}
\begin{enumerate}
\item For each joint event, the access distance (sum) is calculated from the initial anomaly service(s) to the service where the event belongs to. If any ``access'' is not reachable, the distance is set as $d_m+1$ where $d_m$ is the maximum possible distance. The one with shorter access distance (sum) would be ranked higher and vice versa. Figure \ref{fig:untie} presents an example, where \emph{Service-A} and \emph{Service-B} are both initial anomaly services. Since \system suspects that $event 2$ is caused by either $event 3$ or $event 1$ with the same weight. The scores of $event 3$ and $event 1$ are tied. Then, $event 3$ has a score of $1$ (i.e., $0+1$) and $event 1$ has a score of 2 (i.e., $0+2$), since it is not reachable by \emph{Service-B}). Therefore, $event 3$ is ranked first and logical. 
\item For the remaining joint results with the same access distances, \system continues to untie by using the historical root cause frequency of the event types under the same trigger conditions (e.g., checkout domain alerts). This frequency information is generated from the manually labeled dataset. A more frequently occurred root cause type is ranked higher.
\end{enumerate}

\subsection{Rule Customization Management}

While \system users create or update the rules,  there could be overlaps, inconsistencies, or even conflicts being introduced such as the example in Figure~\ref{fig:ex2_n2}. \system uses two graphs to manage the rule relationships and avoid conflicts for users. One graph is to represent the link rules between events in the same service (\emph{Same-Graph}) while the other is to represent links between different services (\emph{Diff-Graph}). The nodes in these two graphs are the event types defined in Section~\ref{sec:causality}. There are three statuses between each (directional) pair of event types: (1) no rule, (2) only basic rule, and (3) conditional rule (since it overwrites the basic rule). In \emph{Same-Graph}, \system does not allow self-loop as it does not build links between an event and itself.

When rule change happens, existing rules are enumerated to build edges in \emph{Same-Graph} and \emph{Diff-Graph} based on $Target\mbox{-}Events$ and $Target\mbox{-}Service$. Based on the users' operation of 
    (1) ``remove a rule'',  \system removes the corresponding edge on the graphs;
    (2) ``add/update a rule'',  \system checks whether there are existing edges between the given event types, and then warns the users for possible overwrites. 
    If there are no conflicts, \system just adds/updates edges between the event types.

After all changes, \system extracts the rules from the graphs by converting each edge to a single rule. These rules are automatically implemented, and then tested against our labeled data set. The \system users need to review the changes with validation reports before the changes go online.

\section{Evaluation}

We evaluate \system in two aspects: (1) \emph{effectiveness (accuracy)}, which assesses how accurate \system is in  detecting and ranking root causes, and (2) \emph{efficiency}, which assesses how long it takes for \system to derive root causes and conduct end-to-end analysis in action. Particularly, we intend to address the following research questions:

\begin{itemize}
    \item \textbf{RQ1.} What are the accuracy and efficiency of \system when applied on the collected dataset?
    \item \textbf{RQ2.} How does \system compare with baseline approaches in terms of accuracy?
    \item \textbf{RQ3.} What are the accuracy and efficiency of \system in an end-to-end scenario?
\end{itemize}


\subsection{Evaluation Setup}
\label{sec:evalset}
To evaluate \system in a real-world scenario, we deploy and apply \system  in eBay's e-commerce system that serves more than 159 million active buyers. In particular, we apply \system upon a microservice ecosystem that contains over 5,000 services on three data centers. These services are built on different tech stacks with different programming languages, including Java, Python, Node.js, etc. Furthermore, these services interact with each other by using different types of service protocols, including HTTP, gRPC,  and Message Queue. The distributed tracing of the ecosystem generates 147B traces on average per day.
\subsubsection{Data Set}
\label{sec:dataset}
The SRE teams at eBay help collect a labeled data set containing 952 incidents over 15 months (Jan 2020 - Apr 2021). Each incident data contains the input required by \system (e.g., dependency snapshot and events with details) and the root cause manually labeled by the SRE teams. 
These incidents are grouped into two categories: 
\begin{itemize}
\item \emph{Business domain incidents.} These incidents are detected mainly due to their business impact. For example, end users encounter failed interactions, and business or customer experience is impacted, similar to the example in  Figure~\ref{fig:example1}. 
\item \emph{Service-based incidents.} These incidents are detected mainly due to their impact on the service level, similar to the example in Figure~\ref{fig:dynamic_example}.
\end{itemize}

An internal incident may get detected early, and then likely get categorized as a service-based incident or even solved directly by owners without records. On the other hand, infrastructure-level issues or issues of external service providers (e.g., checkout and shipping services) may not get detected until business impact is caused. 

There are 782 business domain incidents and 170 service-based incidents in the data set. For each incident, the root cause is manually labeled, validated, and collected by the SRE teams, who handle the site incidents everyday. For a case with multiple interacting causes, only the most actionable/influential event is labelled as the root cause for the case. These actual root causes and incident contexts serve as the ground truth in our evaluation.

\subsubsection{\system Setup}

The \system production system is deployed as three microservices and federated in three data centers with nine 8-core CPUs, 20GB RAM pods each on Kubernetes.


\subsubsection{Baseline Approaches} 
\label{sec:baseline}
In order to compare \system with other related approaches, we design and implement two baseline approaches for the evaluation:  
\begin{itemize}
    \item \emph{Naive Approach.} This approach directly uses the constructed service dependency graph (Section~\ref{sec:appgraph}). The events are assigned a score by the severeness of the associated anomaly. Then a normalized score for each service is calculated summarizing all the events related to the service. Lastly, the PageRank algorithm is used to calculate the root cause ranking. 
    \item \emph{Non-adaptive Approach.} This approach is not context-aware. It replaces all special rules (i.e., conditional and dynamic ones) with their basic rule versions. Its other parts are identical to \system.
\end{itemize}
The non-adaptive approach can be seen as a baseline for reflecting a group of graph-based approaches (e.g.,  CauseInfer~\cite{chen2014causeinfer} and Microscope~\cite{lin2018microscope}). These approaches also specify certain service-level metrics but lack the context-aware capabilities of \system. Because the tools for these approaches are not publicly available, we implement the non-adaptive approach to approximate these approaches.

\subsection{Evaluation Results}


\subsubsection{RQ1}

\label{sec:rq1}

\begin{table}[t]
\centering
\caption{Accuracy of RCA by \system and baselines}
\resizebox{0.9\linewidth}{!}{ 
\begin{tabular}{l|r|r|r|r|r|r|}
\cline{2-7}
                                    & \multicolumn{2}{c|}{\system} & \multicolumn{2}{c|}{Naive} & \multicolumn{2}{c|}{Non-adaptive} \\ \cline{2-7} 
                                    & Top 3        & Top 1       & Top 3         & Top 1      & Top 3         & Top 1 \\ \hline
\multicolumn{1}{|c|}{Service-based}    & 92\%         & 74\%         & 25\%          & 16\%   & 84\% & 62\%       \\ \hline
\multicolumn{1}{|c|}{Business domain} & 96\%         & 81\%        & 2\%          & 1\%   & 28\% & 26\%       \\ \hline
\multicolumn{1}{|c|}{Combined} & 95\%         & 78\%        & 6\%          & 3\%   & 38\% & 33\%       \\ \hline
\end{tabular}
}
 \label{tab:accuracy}
  \vspace{-3.0ex} 
\end{table}


Table~\ref{tab:accuracy} shows the results of applying \system on the collected data set. We measure both top-1 and top-3 accuracy. The top-1 and top-3 accuracy is calculated as the percentage of cases where their ground-truth root cause is ranked within top 1 and top 3, respectively, in \system's results. 
\system achieves high accuracy on both incident categories. For example, for business domain incidents, \system achieves 96\% top-3 accuracy.

The unsuccessful cases that \system ranks the root cause after top 3 are mostly caused by missing event(s). 
More than one-third of these unsuccessful cases have been addressed by adding necessary events and corresponding rules over time. For example, initially, we had only an event type of general error spike, which mixes different categories of errors and thus causes high false-positive rate. We then have designed different event types for each category of the error metrics (including various internal and client API errors). 
In many cases that \system ranks the root cause after top 1, the labeled root cause is just one of the multiple co-existing root causes. But for fairness,  the SRE teams label only a single root cause in each case. According to the feedback from the SRE teams, \system still facilitates the RCA process for these cases.   

Our results show that the runtime cost of applying \system is relatively low. For a service-based incident, the average runtime cost of \system is 1.06s while the maximum is 1.69s. For a business domain incident, the average runtime cost is 0.98s while the maximum is 1.14s. 


\subsubsection{RQ2}

\label{sec:rq2}

We additionally apply the baseline approaches on the  data set. Table~\ref{tab:accuracy} also shows the evaluation results. The results show that the accuracy of \system is substantially higher than that of the baseline approaches. In terms of the top-1 accuracy, \system achieves 78\% compared with 3\% and 33\% of the naive and non-adaptive approaches, respectively.  In terms of the top-3 accuracy, \system achieves 95\% compared with 6\% and 38\% of the naive and non-adaptive approaches, respectively. 

The naive approach performs worst in all settings, because it blindly propagates the score at service levels.
The accuracy of the non-adaptive approach is much worse for business domain incidents. The reason is that for a business domain incident, it often takes a longer propagation path since the incident is triggered by a group of services, and new dynamic dependencies may be introduced during the event collection, causing more inaccuracy for the non-adaptive approach. 
There can be many non-critical or irrelevant error events in an actual production scenario, aka ``soft'' errors. We suspect that these non-critical or irrelevant events may be ranked higher by the non-adaptive approach since they are similar to injected faults and hard to be distinguished from the actual ones. \system uses dynamic and conditional rules to discover the actual causal links, building fewer links related to such non-critical or irrelevant events for leading to higher accuracy.  

\subsubsection{RQ3}

\begin{table}[]
\centering
\caption{Comparison of \system results on the dataset and end-to-end scenario}
\resizebox{0.95\linewidth}{!}{ 
\begin{tabular}{l|r|r|r|r|}
\cline{2-5}
                                             & \multicolumn{2}{c|}{Service-based} & \multicolumn{2}{c|}{Business Domain} \\ \cline{2-5} 
                                             & Dataset          & End-to-End          & Dataset         & End-to-End         \\ \hline
\multicolumn{1}{|l|}{Top-1 Accuracy}         & 74\%             & 73\%                & 81\%            & 73\%               \\ \hline
\multicolumn{1}{|l|}{Top-3 Accuracy}         & 92\%              & 91\%                & 96\%            & 87\%               \\ \hline
\multicolumn{1}{|l|}{Average Runtime Cost} & 1.06s            & 3.16s               & 0.98s           & 2.98s              \\ \hline
\multicolumn{1}{|l|}{Maximum Runtime Cost} & 1.69s            & 4.56s               & 1.14s           & 3.61s              \\ \hline
\end{tabular}
}
\label{tab:compare}
 \vspace{-3.0ex} 
\end{table}

To evaluate \system under an end-to-end scenario, we apply \system  upon actual incidents in action. Table~\ref{tab:compare} shows the results. The accuracy has a decrease of up to 9 percentage points in the end-to-end scenario, with some failures caused by production issues such as missing data and service/storage failures. In addition, the runtime cost is increased by up to nearly 3 seconds due to the time spent on fetching data from different data sources, e.g., querying the events for a certain time period.

\section{Experience}

\begin{figure}[t]
\centering
  \includegraphics[width=0.88\linewidth]{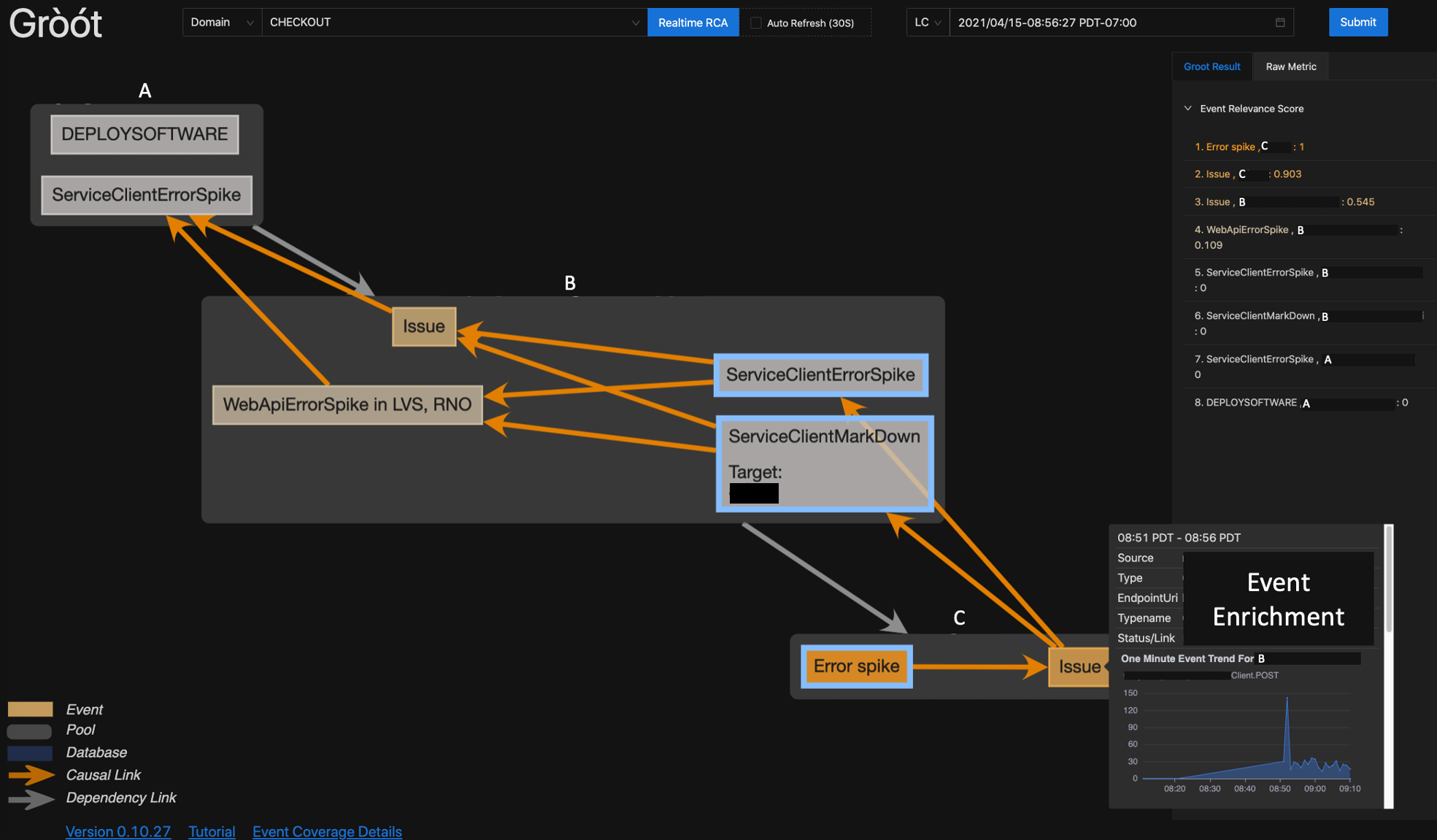}
  \caption{\system UI in production}
  \label{fig:UI}
   \vspace{-3.0ex} 
\end{figure}

\system currently supports daily SRE work. Figure~\ref{fig:UI} shows a live \system's ``bird's eye view'' UI on an actual simple checkout incident. Service $C$ has the root cause ($Error Spike$) and belongs to an external provider. Although the domain service $A$ also carries an error spike and gets impacted, \system correctly ignores the irrelevant deployment event, which has no critical impact. The events on $C$ are virtually created based on the dynamic rule. 
Note that all causal links (yellow) in the UI indicate ``is  cause of'', being the opposite of ``is caused by'' as  described in Section~\ref{sec:causality} to provide more intuitive UI  for users to navigate through. \system visualizes the dependency and event causality graph with extra information such as an error message. The SRE teams can quickly comprehend the incident context and derived root cause to investigate \system further. A mouseover can trigger ``event enrichment'' based on the event type to present details such as raw metrics and other additional information.

We next share two major kinds of experience:
\begin{itemize}
   \item \textbf{Feedback from \system users and developers},  reflecting the general experience of two groups: (1) domain SRE teams who use \system to find the root cause, and 
    (2) a centered SRE team who maintains \system to facilitate new requirements. 
   \item \textbf{Lessons learned}, representing the lessons learned from deploying and adopting \system in production for the real-world RCA process.
\end{itemize}

\subsection{Feedback from \system Users and Developers}

We invite the SRE members who use \system for RCA in their daily work to the user survey. We call them users in this section. We also invite different SRE members responsible for maintaining \system to the developer survey. We call them developers in this section. In total, there are 14 users and 6 developers\footnote{The \system researchers and developers who are authors of this paper are excluded.} who respond to the surveys. 

For the user survey, we ask 14 users the following 5 questions (Questions 4-5 have the same choices as Question 1):
\begin{itemize}
    \item \textbf{Question 1.} When \system correctly locates the root cause, how does it help with your triaging experience? Answer choices: Helpful(4), Somewhat Helpful(3), Not Helpful(2), Misleading(1).
    \item \textbf{Question 2.} When \system correctly locates the root cause, how does it save/extend your or the team's triaging time? (Detection and remediation time not included) Answer choices: Lots Of Time Saved(4), Some Time Saved(3), No Time Saved(2), Waste Time Instead(1).
    \item \textbf{Question 3.} Based on your estimation, how much triage time \system would save on average when it correctly locates the root cause? (Detection and remediation time  not included) Answer choices: More than 50\%(4), 25-50\%(3), 10-25\%(2), 0-10\%(1), N/A(0).
    \item \textbf{Question 4.} When \system correctly locates the root cause, do you find that the result ``graph'' provided by \system helps you understand how and why the incident happens?
    \item \textbf{Question 5.} When \system does not correctly locate the root cause, does the result ``graph'' make it easier for your investigation of the root cause?
\end{itemize}

\begin{figure}
    \centering
    \begin{subfigure}[t]{0.45\linewidth}
    \begin{tikzpicture}[scale=0.45]
  \begin{axis}
    [
    ytick={1,2,3,4,5},
    yticklabels={Question 5, Question 4, Question 3, Question 2, Question 1},
    xmin=0
    ]
    \addplot+[
    boxplot prepared={
      average=3.43,
      upper quartile=4.0,
      lower quartile=3.0,
      upper whisker=4.0,
      lower whisker=2.0
    },
    ] coordinates {};
    \addplot+[
    boxplot prepared={
      average=3.64,
      upper quartile=4.0,
      lower quartile=3.0,
      upper whisker=4.0,
      lower whisker=3.0
    },
    ] coordinates {};
    \addplot+[
    boxplot prepared={
      average=2.79,
      upper quartile=3.25,
      lower quartile=2.0,
      upper whisker=4.0,
      lower whisker=2.0
    },
    ] coordinates {};
    \addplot+[
    boxplot prepared={
      average=3.71,
      upper quartile=4.0,
      lower quartile=3.0,
      upper whisker=4.0,
      lower whisker=3.0
    },
    ] coordinates {};
    \addplot+[
    boxplot prepared={
      average=3.79,
      upper quartile=4,
      lower quartile=3.75,
      upper whisker=4.0,
      lower whisker=3.0
    },
    ] coordinates {};
  \end{axis}
\end{tikzpicture}
\caption{From 14 \system users}
\label{fig:survey1}
\end{subfigure}
~
\begin{subfigure}[t]{0.45\linewidth}
    \begin{tikzpicture}[scale=0.45]
  \begin{axis}
    [
    ytick={1,2,3,4,5},
    yticklabels={Question 5, Question 4, Question 3, Question 2, Question 1},
    xmin=0
    ]
    \addplot+[
    boxplot prepared={
      average=3.83,
      upper quartile=4,
      lower quartile=3.75,
      upper whisker=4.0,
      lower whisker=3.0
    },
    ] coordinates {};
    \addplot+[
    boxplot prepared={
      average=3.5,
      upper quartile=4.0,
      lower quartile=3.0,
      upper whisker=4.0,
      lower whisker=3.0
    },
    ] coordinates {};
    \addplot+[
    boxplot prepared={
      average=3.67,
      upper quartile=4.0,
      lower quartile=3.0,
      upper whisker=4.0,
      lower whisker=3.0
    },
    ] coordinates {};
    \addplot+[
    boxplot prepared={
      average=3.67,
      upper quartile=4.0,
      lower quartile=3.0,
      upper whisker=4.0,
      lower whisker=3.0
    },
    ] coordinates {};
    \addplot+[
    boxplot prepared={
      average=3.83,
      upper quartile=4,
      lower quartile=3.75,
      upper whisker=4.0,
      lower whisker=3.0
    },
    ] coordinates {};
  \end{axis}
\end{tikzpicture}
\caption{From 6 \system developers}
\label{fig:survey2}
\end{subfigure}
\caption{Survey results}
 \vspace{-3.0ex} 
\end{figure}
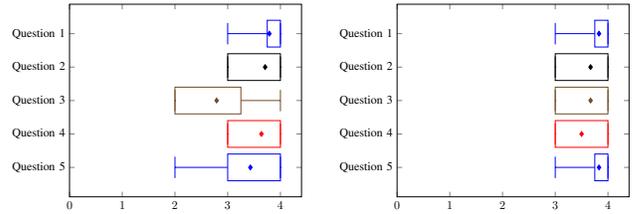

Figure~\ref{fig:survey1} shows the results of the user survey. We can see that most users find \system very useful to locate the root cause. The average score for Question 1 is 3.79, and 11 out of 14 participants find \system very helpful. As for Question 3, \system saves the triage time by 25-50\%. Even in cases that \system cannot correctly locate the root cause, it is still helpful to provide information for further investigation with an average score of 3.43 in Question 5.

For the developer survey, we ask the 6 developers the following 5 questions (Questions 2-5 have the same choices as Question 1):
\begin{itemize}
    \item \textbf{Question 1.} Overall, how convenient is it to change and customize events/rules/domains while using  \system? Answer choices: Convenient(4), Somewhat Convenient(3), Not Convenient(2), Difficult(1).
    \item \textbf{Question 2.} How convenient is it to \emph{change/customize event models} while using \system? 
    \item \textbf{Question 3.} How convenient is it to \emph{add new domains} while using \system? 
    \item \textbf{Question 4.} How convenient is it to \emph{change/customize causality rules} while using \system? 
    \item \textbf{Question 5.} How convenient is it to change/customize \system compared to other SRE tools? 
\end{itemize}

Figure~\ref{fig:survey2} shows the results of the developer survey. Overall, most developers find it convenient to make changes on and customize events/rules/domains in \system. 

\subsection{Lessons learned}
In this section, we share the lessons learned in terms of technology transfer and adoption on using \system in production environments.

\emph{Embedded in Practice.} To build a successful RCA tool in practice, it is important to embed the R\&D efforts in the live environment with SRE experts and users. We have a 30-minute routine meeting daily with an SRE team to manually test and review every site incident. In addition, we actively reach out to the end users for feedback. For example, the users found our initial UI hard to understand. Based on their suggestions, we have introduced alert enrichment with the detailed context of most events, raw metrics, and links to other tools for the next steps. We also make the UI interactive and build user guides, training videos, and sections. As a result, \system has become increasingly practical and well adopted in practice. We believe that R\&D work on observability should be incubated and grown within daily SRE environments. It is also vital to bring developers with rich RCA experience into the R\&D team.

\emph{Vertical Enhancements.} High-confidence and automated vertical enhancements can empower great experiences. \system is enhanced and specialized in critical scenarios such as grouped related 
alerts across services or critical business domain issues, and large-scale scenarios such as infrastructure changes or database issues. Furthermore, the end-to-end automation is also built for integration and efficiency with anomaly detection, RCA, and notification. For notification, domain business anomalies and diagnostic results are sent through communication apps (e.g., slack and email) for better reachability and experience. Within 18 months of R\&D, \system now supports 18 business domains and sub-domains of the company. On average, \system UI supports more than 50 active internal users, and the service sends thousands of results every month. Most of these usages are around the vertical enhancements. 


\emph{Data and Tool Reliability.} Reliability is critical to \system itself and requires a lot of attention and effort. For example, if a critical event is missing, \system may infer a totally different root cause, which would mislead users. 
We estimate the alert accuracy 
to be greater than 0.6 in order to be useful. Recall is even more important since \system can effectively eliminate false positive alerts based on the casual ranking. Since there are hundreds of different metrics supported in \system, we spend time to ensure a robust back end by adding partial and dynamic retry logic and high-efficiency cache. \system's unsuccessful cases can be caused by imperfect data, flawed algorithms, or simply code defects. To better trace the reason behind each unsuccessful case, we add a tracing component. Every \system request can be traced back to atomic actions such as retrieving data, data cleaning, and anomaly detection via algorithms.

\emph{Trade-off among Models.} The accuracy and scalability trade-off among anomaly detection models should be carefully considered and tested. In general, some algorithms such as deep-learning-based or ensemble models are more adaptive and accurate than typical ones such as traditional ML or statistical models. However, the former requires more computation resources, operational efforts, and additional system complexities such as training or model fine-tuning. Due to the actual complexities and fast-evolving nature of our context, it is not possible to scale each model (e.g., deep-learning-based models), nor have it deeply customized for every metric at every level. Therefore, while selecting models, we must make careful trade-off in aspects such as accuracy, scalability, efficiency, effort, and robustness. In general, we first set different ``acceptance'' levels by analyzing each event's impact and frequency, and then test different models in staging and pick the one that is good enough. For example, a few alerts such as ``high thread usage'' are defined by thresholds and work just fine even without a model. Some alerts such as  ``service client error'' are more stochastic and require coverage on every metric of every service, and thus we select fast and robust statistical models and actively conduct detection on the fly.  

\emph{Phased Incorporation of ML.} In the current industrial settings, ML-powered RCA products still require effective knowledge engineering. Due to the higher complexity and lower ``signal to noise ratio'' of real production incidents, many existing approaches cannot be applied in practice. We believe that the knowledge engineering capabilities can facilitate adoption of technologies such as AIOps. Therefore, \system is designed to be highly customizable and easy to infuse SRE knowledge and to achieve high effectiveness and efficiency. Moreover, a multi-scenario RCA tool requires various and interpretable events from different detection strategies. Auto-ML-based anomaly detection or unsupervised RCA for large service ecosystems is not yet ready in such context.
As for the path of supervised learning, the training data is tricky to label and vulnerable to potential cognitive bias. 
Lastly, the end users often require complete understanding to fully adopt new solutions, because there is no guarantee of correctness. Many recent ML algorithms (e.g., ensemble and deep learning) lack interpretability. Via the knowledge engineering and graph capabilities, \system is able to explain diversity and causality between ML-model-driven and other types of events. Moving forward, we are building a white-box deep learning approach with causal graph algorithms where the causal link weights are parameters and derivable. 

\section{Conclusion}

In this paper, we have presented our work around root cause analysis (RCA) in industrial settings. To tackle three major RCA challenges (complexities of operation, system scale, and monitoring), we have proposed a novel event-graph-based approach named  \system that constructs a real-time causality graph for allowing adaptive customization. 
\system can handle diversified anomalies and activities from the system under analysis and is extensible to different approaches of anomaly detection or RCA. We have integrated \system into eBay's large-scale distributed system  containing more than \textbf{5,000} microservices. Our evaluation of \system on a data set consisting of 952 real production incidents shows that \system achieves high accuracy and efficiency across different scenarios and also largely outperforms  baseline graph-based approaches. We also share the lessons learned from deploying and adopting \system in production environments.

\bibliographystyle{IEEEtran}
\bibliography{references}

\end{document}